\documentclass[11pt]{article}
\usepackage[margin=1in]{geometry}
\usepackage{amsmath,amssymb}
\usepackage{booktabs}
\usepackage{array}
\usepackage{threeparttable}
\usepackage{multirow}
\usepackage{graphicx}
\usepackage{placeins}
\usepackage[round]{natbib}
\usepackage{hyperref}
\usepackage{xurl}
\hypersetup{hidelinks}
\newcommand{\E}{\mathbb{E}}
\newcommand{\Qbar}{\bar{Q}}
\DeclareMathOperator{\expit}{expit}
\newcommand{\indep}{\perp\!\!\!\perp}
\graphicspath{{figures/}}

\begin{document}

\title{Improving Efficiency and Power in Small Rare-Disease Trials Through Prespecified, Data-Adaptive Covariate Adjustment}

\author{%
  Tianyue Zhou\textsuperscript{1}\thanks{Corresponding author. Email: \nolinkurl{tianyue_zhou@berkeley.edu}},
  Susan Gruber\textsuperscript{2}, Hana Lee\textsuperscript{3},\\
  Wonyul Lee\textsuperscript{3}, Lei Nie\textsuperscript{3},
  and Mark van der Laan\textsuperscript{1}\\[0.75em]
  \textsuperscript{1}Division of Biostatistics, School of Public Health,\\
  University of California, Berkeley, CA, USA\\
  \textsuperscript{2}Targeted ML Solutions, Cambridge, MA, USA\\
  \textsuperscript{3}Center for Drug Evaluation and Research,\\
  U.S. Food and Drug Administration (FDA), Silver Spring, MD, USA%
}

\date{}

\maketitle

\begin{abstract}
Covariate adjustment can potentially improve efficiency in randomized clinical trials, but in rare diseases, limited prior knowledge makes it difficult to specify an adjustment at the design stage. Targeted maximum likelihood estimation (TMLE) with the Super Learner (TMLE+SL) can outperform fixed adjustment methods, such as the G-computation estimator, by leveraging machine learning to learn both the covariates and the functional form of the adjustment from the trial data. Estimating the known treatment mechanism can improve efficiency further. In simulations reflecting real rare-disease trials, TMLE+SL yielded valid inference and achieved greater efficiency and power than conventional analyses. Finally, we also provide practical guidance on specifying TMLE+SL in rare-disease trials.
\end{abstract}


\section{Introduction}

Regulatory guidance explicitly recognizes covariate adjustment as a tool for improving statistical efficiency in randomized clinical trials (RCTs). The 2023 guidance of the U.S. Food and Drug Administration (FDA) centers this recommendation on prognostic baseline covariates, emphasizing that adjusting for them can improve the precision of treatment-effect estimates and the power of the corresponding tests \citep{fda2023covariate}. The guideline of the European Medicines Agency makes similar recommendations \citep{ema2015}. Implementing covariate adjustment in a confirmatory trial requires commitment before unblinding: the statistical analysis plan specifies both the covariates to be adjusted for and the analysis method \citep{fda2023covariate}. The benefit of adjustment, however, depends on how prognostic the chosen covariates are: adjusting for variables only weakly related, or unrelated, to the endpoint provides little gain and can even cost precision \citep{fda2023covariate}. Moreover, the properties of adjustment are best understood when the number of covariates is small relative to the sample size \citep{tsiatis2008,fda2023covariate}.

In rare-disease trials, these two choices, which covariates to adjust for and in what form, are especially difficult to make in advance. Such trials typically enroll few participants, who nevertheless vary widely in disease stage, phenotype, genotype, and endpoint trajectory, while the natural history of the disease is often incompletely characterized and disease-specific evidence remains limited \citep{fda2023rare,chen2025rarechallenges}. As a result, many baseline variables may be clinically credible predictors of the primary endpoint, yet the evidence available at the design stage is often insufficient to rank their prognostic value or to specify how they relate to the endpoint, and the small sample limits the number of covariates that a single outcome model can support. Under these conditions, a prespecified adjustment, in both its covariates and their form, may not match the prognostic structure of the data, and any mismatch costs efficiency and power. Sponsors may therefore prefer a conservative compromise, such as adjusting only for the stratification factors and the baseline value of the endpoint, and accept some potential loss of efficiency. Procedures that are fully specified in the statistical analysis plan, yet estimate the strength and form of the covariate--outcome relationships from the trial data, would remove the need for this compromise.

Targeted learning provides one such procedure. Targeted maximum likelihood estimation (TMLE) is a general template for constructing regular and asymptotically efficient substitution estimators of a statistical estimand \citep{vdl2006tmll,vdlrose2011}. It can incorporate machine-learning algorithms to reduce the risk of model misspecification while retaining valid statistical inference, typically through the Super Learner (SL), which uses cross-validation either to select the best-performing estimator from a prespecified library of candidates or to form their optimal weighted combination \citep{vdl2007superlearner}. Within the extensive literature on targeted learning for RCTs \citep{moore2009,rosenblum2010simple,moore2011robust}, the work most relevant to this paper is that of \citet{balzer2016}, who built on empirical efficiency maximization \citep{rubin2008eem} to select the candidate estimator with the smallest cross-validated variance, thereby targeting precision, and hence power, directly. Their approach was later extended to richer libraries that include machine-learning algorithms \citep{balzer2024adaptive,balzer2026journey}. Even though the treatment mechanism is known by design in an RCT, estimating it from the baseline covariates can yield further efficiency gains \citep{vdlrobins2003,balzer2016}. TMLE has served as the primary analysis of the SEARCH community-randomized HIV test-and-treat trial \citep{havlir2019search} and has been used in regulatory submissions, as documented in a recent landscape analysis \citep{tmleLandscape}. However, how this approach performs in rare-disease trials, and how it should be specified for them, have received comparatively little attention.

This paper addresses both questions. First, to relate the approach to current practice, we show that some of the covariate-adjustment methods recommended in the FDA guidance can be viewed as special cases of TMLE. Second, we illustrate how efficiency and power can be improved by fully exploiting the flexibility of TMLE+SL, as we refer to the procedure of \citet{balzer2016}, and we evaluate its performance in a simulation study calibrated to a completed rare-disease trial. Third, we offer practical guidance on specifying TMLE+SL in such trials. We focus on the simplest confirmatory setting, a point treatment with a fully observed continuous endpoint, because it is common in rare-disease trials and raises specification choices that more complex designs share. Section~\ref{sec:stat_problem} defines the estimation problem, and Section~\ref{sec:tmle_overview} reviews the TMLE of the average treatment effect in an RCT. Section~\ref{sec:gcomp_tmle} shows that the G-computation estimator described in the guidance, when its working model is a standard parametric regression, is itself a TMLE. Section~\ref{sec:tmle_sl_power} describes how TMLE+SL can improve on this special case by data-adaptively learning the outcome regression and estimating the known treatment mechanism, Section~\ref{sec:simulation} presents the simulation study, and Section~\ref{sec:conclusion} concludes.

\section{Defining the statistical estimation problem}\label{sec:stat_problem}

\subsection{Observed data}\label{subsec:observed_data}

Consider a two-arm RCT enrolling $n$ participants. For each participant $i=1,\dots,n$, we observe
\begin{equation}\label{eq:observed_data}
O_i = (W_i,A_i,Y_i),
\end{equation}
where $W_i$ denotes baseline covariates, $A_i\in\{0,1\}$ denotes the randomized binary treatment assignment, and $Y_i$ denotes the outcome. We assume the observations are independent and identically distributed according to an unknown distribution $P_0$ in a nonparametric statistical model $\mathcal{M}$. The treatment mechanism
\begin{equation}\label{eq:known_g}
g_0(a \mid W) = \mathbb{P}_{P_0}(A=a \mid W), \qquad a\in\{0,1\},
\end{equation}
is known by trial design. For simplicity, we assume that the outcome is fully observed for every participant. In practice, some outcomes may be missing, for example because of loss to follow-up. TMLE handles this situation by additionally estimating the missingness mechanism \citep{gruber2012tmle,gruber2023sap}, an extension that this paper does not pursue.

\subsection{Causal estimand, identification, and statistical estimand}\label{subsec:causal_estimand}

Let $Y(1)$ and $Y(0)$ denote the potential outcomes under treatment and control, respectively. The causal estimand is the average treatment effect (ATE),
\begin{equation}\label{eq:causal_ate}
\mathrm{ATE} = \E\!\left\{Y(1)-Y(0)\right\}.
\end{equation}
The randomization assumption, $A \indep \{Y(0),Y(1)\}\mid W$, and the positivity assumption, ${g_0(a\mid W)}>0$ for $a\in\{0,1\}$ and all $W$ in the support, both hold by the trial design. We additionally assume consistency: the observed outcome under treatment $a$ equals the corresponding potential outcome, that is, if $A=a$, then $Y=Y(a)$.

Under these three assumptions, the causal ATE is identified by the statistical estimand
\begin{equation}\label{eq:statistical_estimand}
\psi_0 \equiv \psi(P_0)
  = \E_{P_0}\!\left[\Qbar_0(1,W)-\Qbar_0(0,W)\right],
\end{equation}
where $\Qbar_0(a,w)=\E_{P_0}(Y\mid A=a,W=w)$ is the true outcome regression. In an RCT in which the treatment mechanism does not depend on $W$, $\psi_0$ also equals the difference in mean outcomes between the two arms, $\E_{P_0}(Y\mid A=1)-\E_{P_0}(Y\mid A=0)$ \citep{rosenblum2010simple}.

\subsection{Efficient influence curve}\label{subsec:eic}

For a generic distribution $P\in\mathcal{M}$, not necessarily equal to $P_0$, let $\Qbar(a,w)=\E_P(Y\mid A=a,W=w)$ and $g(a\mid w)=\mathbb{P}_P(A=a\mid W=w)$ denote its outcome regression and treatment mechanism, and let $\psi(P)=\E_P[\Qbar(1,W)-\Qbar(0,W)]$. A subscript $0$ indicates evaluation at $P=P_0$, the true distribution. The efficient influence curve of $\psi$ at $P$ in $\mathcal{M}$ is
\begin{equation}\label{eq:eic}
D^*_P(O)
=
H_g(A,W)\{Y-\Qbar(A,W)\}
+ \Qbar(1,W)-\Qbar(0,W)-\psi(P),
\end{equation}
where $H_g(A,W)=A/g(1\mid W)-(1-A)/g(0\mid W)$. In the targeted learning literature, $H_g$ is termed the \emph{clever covariate} \citep{vdl2006tmll}.

The efficient influence curve characterizes efficient estimation of $\psi_0$: a regular estimator $\widehat{\psi}$ is asymptotically efficient in $\mathcal{M}$ if and only if it is asymptotically linear with influence curve $D^*_{P_0}$ \citep{vdlrobins2003}, that is,
\begin{equation}\label{eq:eic_expansion}
\widehat{\psi}-\psi_0
=
\frac{1}{n}\sum_{i=1}^n D^*_{P_0}(O_i)+o_P\!\left(n^{-1/2}\right),
\qquad\text{hence}\qquad
\sqrt{n}\,(\widehat{\psi}-\psi_0)\;\xrightarrow{d}\;N\!\left(0,\,\mathrm{Var}_{P_0}\{D^*_{P_0}(O)\}\right).
\end{equation}
The expansion~\eqref{eq:eic_expansion} identifies $\mathrm{Var}_{P_0}\{D^*_{P_0}(O)\}$ as the efficiency bound for estimating $\psi_0$, the smallest asymptotic variance attainable by a regular estimator in $\mathcal{M}$ \citep{vdlrobins2003,vdlrose2011}. The TMLE is one estimator constructed to attain this bound under standard regularity conditions.

\section{Targeted maximum likelihood estimation of \texorpdfstring{$\psi_0$}{psi0} in a randomized clinical trial}\label{sec:tmle_overview}

Rather than describing the TMLE procedure in full generality, we specialize it to the estimand $\psi_0$, so that each step can be followed concretely in the RCT setting of this paper. The four steps are initial estimation of the nuisance functions, targeting of the initial fit, point estimation by substitution, and influence-curve-based inference.

\paragraph{Step 1: Initial estimation of the nuisance functions.}
The procedure begins with an initial estimate $\Qbar_n^0(a,w)$ of the true outcome regression $\Qbar_0(a,w)$, which may be produced by any regression method, including flexible machine-learning algorithms such as the Super Learner \citep{vdl2007superlearner}. Although caution is warranted in rare-disease settings, the Super Learner can be used successfully in small samples when its library favors parsimonious parametric working models \citep{balzer2016}. The true treatment mechanism $g_0$, known by design, may be used as is or estimated from the baseline covariates to further improve efficiency. Throughout, $g_n$ denotes whichever of the two the estimator uses.

\paragraph{Step 2: Targeting the initial fit.}
Substituting $\Qbar_n^0$ for $\Qbar_0$ in the identification formula~\eqref{eq:statistical_estimand} and averaging over the empirical distribution of $W$ yields the plug-in estimator $n^{-1}\sum_{i=1}^n\{\Qbar_n^0(1,W_i)-\Qbar_n^0(0,W_i)\}$. For a generic initial estimator, this plug-in estimator does not solve the empirical efficient-influence-curve equation, which sets the sample average of~\eqref{eq:eic}, with $\Qbar$, $g$, and $\psi(P)$ replaced by their estimates, equal to zero. The bias of the plug-in estimator is then not guaranteed to be asymptotically negligible: when $\Qbar_n^0$ is fit by nonparametric methods, it can vanish more slowly than $n^{-1/2}$, so the estimator may fail to satisfy the expansion~\eqref{eq:eic_expansion} and does not, in general, support Wald-type inference. The targeting step therefore updates $\Qbar_n^0$ through a one-dimensional parametric fluctuation submodel, indexed by a parameter $\epsilon$ and constructed from the clever covariate,
\begin{equation}\label{eq:clever_cov}
H_{g_n}(A,W)
=
\frac{A}{g_n(1\mid W)}
-
\frac{1-A}{g_n(0\mid W)},
\end{equation}
such that the score of the submodel at $\epsilon=0$ equals the residual component $H_{g_n}(A,W)\{Y-\Qbar_n^0(A,W)\}$ of the efficient influence curve. Different fluctuation submodels are available. We adopt the linear fluctuation because it reproduces exactly the G-computation estimator that the FDA guidance recommends. This submodel is
\begin{equation}\label{eq:linear_fluct}
\E(Y\mid A,W)=\Qbar_n^0(A,W)+\epsilon\,H_{g_n}(A,W),
\end{equation}
and $\epsilon$ is estimated by the no-intercept least-squares regression of $Y$ on $H_{g_n}$ with the initial fit entered as an offset or, equivalently, by the regression of the residual $Y-\Qbar_n^0(A,W)$ on $H_{g_n}(A,W)$ through the origin. Writing $\widehat{\epsilon}$ for the resulting estimate, we obtain the \emph{targeted} regression by evaluating the update at each treatment level,
\begin{equation}\label{eq:targeted_reg}
\Qbar_n^*(1,\cdot)=\Qbar_n^0(1,\cdot)+\frac{\widehat{\epsilon}}{g_n(1\mid \cdot)},
\qquad
\Qbar_n^*(0,\cdot)=\Qbar_n^0(0,\cdot)-\frac{\widehat{\epsilon}}{g_n(0\mid \cdot)}.
\end{equation}
The least-squares normal equation for $\widehat{\epsilon}$ sets $\sum_{i=1}^n H_{g_n}(A_i,W_i)\{Y_i-\Qbar_n^*(A_i,W_i)\}=0$, that is, the targeted regression solves the residual component of the empirical efficient-influence-curve equation.

\paragraph{Step 3: Point estimation.}
The TMLE of $\psi_0$ is the substitution estimator obtained by evaluating the identification formula~\eqref{eq:statistical_estimand} at the targeted regression $\Qbar_n^*$ and the empirical distribution of $W$,
\begin{equation}\label{eq:tmle_point}
\widehat{\psi}
=
\frac{1}{n}\sum_{i=1}^n
\left\{\Qbar_n^*(1,W_i)-\Qbar_n^*(0,W_i)\right\}.
\end{equation}
The remaining component of the empirical efficient-influence-curve equation, $\sum_{i=1}^n\{\Qbar_n^*(1,W_i)-\Qbar_n^*(0,W_i)-\widehat{\psi}\}$, is then zero by construction, so the TMLE solves this equation in full.

\paragraph{Step 4: Inference.}
Under standard regularity conditions, namely consistent estimation of the outcome regression together with an empirical-process (Donsker) condition, the TMLE is asymptotically linear with influence curve equal to the efficient influence curve $D^*_{P_0}$. For inference, let $\widehat{D}^*(O_i)$ denote the efficient influence curve evaluated at the targeted regression $\Qbar_n^*$, the treatment mechanism $g_n$, and the point estimate $\widehat{\psi}$,
\begin{equation}\label{eq:eic_hat}
\widehat{D}^*(O_i)
=
H_{g_n}(A_i,W_i)\left\{Y_i-\Qbar_n^*(A_i,W_i)\right\}
+\Qbar_n^*(1,W_i)-\Qbar_n^*(0,W_i)-\widehat{\psi}.
\end{equation}
The variance of $\widehat{\psi}$ is estimated by the empirical variance of $\widehat{D}^*$ divided by $n$,
\begin{equation}\label{eq:ic_variance}
\widehat{\sigma}^2
=
\frac{1}{n}\sum_{i=1}^n
\left\{\widehat{D}^*(O_i)-\overline{\widehat{D}^*}\right\}^2,
\qquad
\mathrm{SE}(\widehat{\psi})=\widehat{\sigma}/\sqrt{n},
\end{equation}
where $\overline{\widehat{D}^*}=n^{-1}\sum_{i=1}^n\widehat{D}^*(O_i)$. Wald confidence intervals and tests follow directly. Because the treatment mechanism is known by design or estimated with a parametric working model, the TMLE remains consistent and asymptotically linear even when the initial fit $\Qbar_n^0$ converges to a misspecified limit. In that case, the variance estimator~\eqref{eq:ic_variance} is asymptotically correct when $g_n=g_0$ and conservative when the treatment mechanism is estimated \citep{moore2009,vdlrose2011}. When highly adaptive learners are used for $\Qbar_n^0$, the Donsker condition can be replaced by cross-fitting, as in cross-validated TMLE, in which the initial fit and the remaining steps use separate parts of the sample \citep{zheng2011cvtmle}.

\section{G-computation with parametric working models as a special case of TMLE}\label{sec:gcomp_tmle}

To estimate unconditional treatment effects such as $\psi_0$, the FDA guidance describes the \emph{G-com\-pu\-ta\-tion} estimator, also termed the standardized or plug-in estimator, as one reliable method \citep{fda2023covariate}: an estimate $\Qbar_n^0$ of the outcome regression is used to predict each participant's outcome under treatment and control, and the predicted contrasts are averaged over the trial sample,
\begin{equation}\label{eq:gcomp}
\widehat{\psi}_{\mathrm{gcomp}}
=\frac{1}{n}\sum_{i=1}^n
\left\{\Qbar_n^0(1,W_i)-\Qbar_n^0(0,W_i)\right\},
\end{equation}
which is the plug-in estimator of Step 2 with $\Qbar_n^0$ the fit of a parametric working model. The guidance illustrates the estimator with a logistic model for a binary outcome, where the averaging step is essential because the treatment coefficient of a nonlinear model describes a conditional effect while the estimand is unconditional. The construction itself applies to any outcome type and any working model. For a continuous outcome, the usual choice is the additive linear working model
\begin{equation}\label{eq:linear_wm}
\Qbar_{\beta}(A,W)=\beta_0+\beta_A A+\beta_W^\top S(W),
\end{equation}
where $S(W)$ is a vector of baseline covariates or functions of them, and $\widehat{\beta}$ denotes the ordinary least squares (OLS) estimate of $\beta$. Because the model contains no treatment-by-covariate interaction, the predicted contrast equals $\widehat{\beta}_A$ for every participant, and the G-computation estimator reduces to the analysis of covariance (ANCOVA) estimator $\widehat{\beta}_A$.

The G-computation estimator is itself a TMLE for a broad class of working models, namely generalized linear models with canonical link, fit by maximum likelihood, whose linear predictor contains an intercept and a main term for treatment, as in linear, logistic, and Poisson regression \citep{moore2009,rosenblum2010simple}. To see this, note that when $g_0(1\mid W)$ is constant, the clever covariate $H_{g_0}$ is a linear combination of the intercept and treatment columns of the design matrix. The maximum-likelihood score equations for these two coefficients make the residuals orthogonal to both columns and hence to $H_{g_0}$. The residual component of the empirical efficient-influence-curve equation is thus already zero at the initial fit, and the remaining component is zero for any substitution estimator. The targeting step therefore returns $\widehat{\epsilon}=0$, and the TMLE with this initial fit, the known $g_0$, and the linear fluctuation coincides with the G-computation estimator. Accordingly, the estimator is consistent under arbitrary misspecification of the working model and efficient when the model is correctly specified \citep{rosenblum2010simple}.

In the linear case~\eqref{eq:linear_wm}, the clever covariate takes the form
\begin{equation}\label{eq:clever_span}
H_{g_0}(A,W)=\frac{A}{g_0(1\mid W)}-\frac{1-A}{g_0(0\mid W)}
=-\frac{1}{g_0(0\mid W)}+\frac{A}{g_0(1\mid W)\,g_0(0\mid W)},
\end{equation}
and the OLS normal equations give
\begin{equation}\label{eq:ols_orth}
\sum_{i=1}^n H_{g_0}(A_i,W_i)\left\{Y_i-\Qbar_{\widehat{\beta}}(A_i,W_i)\right\}=0 .
\end{equation}
Hence $\widehat{\epsilon}=0$, $\Qbar_n^*=\Qbar_{\widehat{\beta}}$, and $\widehat{\psi}_{\mathrm{gcomp}}=\widehat{\beta}_A$ is a TMLE.

The correspondence extends to inference. For the linear working model, the estimated efficient influence curve~\eqref{eq:eic_hat} reduces to $\widehat{D}^*(O_i)=H_{g_0}(A_i,W_i)\{Y_i-\Qbar_{\widehat{\beta}}(A_i,W_i)\}$, and its empirical variance estimates the same asymptotic variance of $\widehat{\beta}_A$ as does the Huber--White sandwich estimator. The guidance recommends robust standard errors of this kind for models with no treatment-by-covariate interaction \citep{rosenblum2009regression,fda2023covariate}. For nonlinear working models, the influence-curve variance~\eqref{eq:ic_variance} also provides a standard error with the theoretical justification that the guidance calls for \citep{rosenblum2010simple}.

\section{TMLE+SL for improved efficiency and power}\label{sec:tmle_sl_power}

With a standard parametric working model, the G-computation estimator is consistent and asymptotically linear under arbitrary misspecification (Section~\ref{sec:gcomp_tmle}), but it can be highly inefficient if the model approximates the true outcome regression poorly. Machine learning reduces this risk by estimating the outcome regression more flexibly. However, a plug-in estimator based on a nonparametric fit does not, in general, yield valid confidence intervals and tests. TMLE+SL data-adaptively estimates the outcome regression and the treatment mechanism with the Super Learner \citep{vdl2007superlearner,balzer2016} and then applies the targeting step, which removes the bias and ensures asymptotic linearity. The resulting estimator therefore supports valid inference and is generally more efficient than the G-computation estimator with a single parametric working model.

\subsection{Super Learner for the outcome regression}\label{subsec:sl_qbar}

The Super Learner is an ensemble machine-learning algorithm grounded in loss-based estimation theory \citep{vdl2007superlearner}. Its inputs are a library of candidate prediction algorithms (such as parametric regressions or nonparametric machine-learning methods), a loss function used to evaluate their performance, and a cross-validation scheme. In $V$-fold cross-validation, the observations are partitioned into $V$ folds, each of which serves in turn as the validation fold. Each algorithm is fit on the other $V-1$ folds, and its loss is evaluated on the validation fold. Its cross-validated risk is the average of these losses over all $n$ observations. The Super Learner then returns a regression fit, either the single candidate with the smallest cross-validated risk (the discrete Super Learner) or the weighted combination of candidates that minimizes this risk (the ensemble Super Learner).

All three inputs must be prespecified in the statistical analysis plan, and each requires particular care. First, the loss function should be chosen according to the objective of the analysis, which here is to improve efficiency and power in estimating $\psi_0$. Following the principle of empirical efficiency maximization \citep{rubin2008eem,balzer2016}, we therefore adopt the squared efficient influence curve ($\mathrm{eIC}^2$) evaluated at the known $g_0$ as the loss function for each candidate. The risk then corresponds to the asymptotic variance of the TMLE based on that candidate, and its cross-validated estimate is
\begin{equation}\label{eq:cv_eic}
\widehat{R}_{\mathrm{CV}}
=\frac{1}{n}\sum_{i=1}^n
\Big[
H_{g_0}(A_i,W_i)\big\{Y_i-\Qbar_{n,-v(i)}(A_i,W_i)\big\}
+\Qbar_{n,-v(i)}(1,W_i)-\Qbar_{n,-v(i)}(0,W_i)-\widehat{\psi}_{v(i)}
\Big]^2,
\end{equation}
where $v(i)$ denotes the fold containing observation $i$, $\Qbar_{n,-v}$ the candidate's fit on the observations outside fold $v$, and $\widehat{\psi}_{v}=n_v^{-1}\sum_{i:\,v(i)=v}\{\Qbar_{n,-v}(1,W_i)-\Qbar_{n,-v}(0,W_i)\}$ the plug-in estimate on the $n_v$ observations of fold $v$. With this loss, the Super Learner selects the candidate that minimizes~\eqref{eq:cv_eic}, directly targeting precision and power.

Second, in small rare-disease trials, the library should consist mainly of parametric working models. It should include at least the unadjusted estimator and generalized linear models with clinically plausible adjustment sets. By the oracle inequality for the cross-validation selector \citep{vdl2003unified,vdl2007superlearner}, TMLE+SL is then no less efficient than any of them. In particular, it matches the covariate-adjusted analysis that would otherwise have served as primary when that analysis is the best candidate, and it improves on that analysis when a better candidate is in the library. Least absolute shrinkage and selection operator (LASSO) regression \citep{tibshirani1996} or simple splines may be added for greater flexibility. More adaptive learners, such as shallow, well-regularized gradient-boosted trees, can sometimes perform well even in small samples and may also be considered in the library.

Third, the number of folds $V$ should be large, so that each candidate is fit on nearly all $n$ observations. Leave-one-out or $20$-fold cross-validation is often suitable \citep{phillips2023practical}. When $V<n$, the statistical analysis plan should also fix how the folds are formed, including the random seed and the ordering of participants before splitting, so that the folds cannot be chosen after the data are seen.

\subsection{Super Learner for the treatment mechanism}\label{subsec:sl_g}

Even when the outcome regression is estimated with the Super Learner, the limited degrees of freedom available in a small sample may prevent the fit from fully capturing the true covariate--outcome relationship. For example, it may not be possible to include all prognostic covariates or to represent nonlinear relationships. In this case, targeting with an estimate of the known treatment mechanism provides additional degrees of freedom to recover part of what the fitted outcome regression misses and can thereby further improve efficiency \citep{vdlrobins2003,balzer2016}.

Given the fitted outcome regression, TMLE+SL evaluates the candidate estimators of the treatment mechanism with the same loss function. Algorithm~\ref{alg:tmle_sl} summarizes the complete procedure.

\newcounter{algorithm}
\begin{figure}[htbp]
\refstepcounter{algorithm}\label{alg:tmle_sl}
\small
\setlength{\defaultaddspace}{5pt}
\begin{tabular}{@{}p{\linewidth}@{}}
\toprule
\textbf{Algorithm~\thealgorithm}\enspace TMLE+SL\\
\midrule
\textbf{Inputs.} A library of $K$ candidate estimators of the outcome regression, a library of $J$ candidate estimators of the treatment mechanism, and a $V$-fold cross-validation scheme, all prespecified.\\
\addlinespace
\textbf{Step 1: Initial estimation of the nuisance functions.}\\
\hangindent=1.75em \hangafter=1 \makebox[1.75em][l]{(a)}For each candidate estimator $k=1,\dots,K$ of the outcome regression, let $\Qbar^{(k)}_{n,-v}$ be its fit on the observations outside fold $v$. The Super Learner selects the candidate $\widehat{k}$ that minimizes~\eqref{eq:cv_eic} evaluated at $\Qbar^{(k)}_{n,-v}$, and $\Qbar_n^0$ is its fit on all $n$ observations.\\
\hangindent=1.75em \hangafter=1 \makebox[1.75em][l]{(b)}For each candidate estimator $j=1,\dots,J$ of the treatment mechanism, let $g^{(j)}_{n,-v}$ be its fit on the observations outside fold $v$. On the same observations, fitting~\eqref{eq:linear_fluct} with offset $\Qbar^{(\widehat{k})}_{n,-v}$ and clever covariate $H_{g^{(j)}_{n,-v}}$ gives $\widehat{\epsilon}^{(j)}_{-v}$ and the targeted regression $\Qbar^{*(j)}_{n,-v}=\Qbar^{(\widehat{k})}_{n,-v}+\widehat{\epsilon}^{(j)}_{-v}H_{g^{(j)}_{n,-v}}$. The Super Learner selects the candidate $\widehat{j}$ that minimizes~\eqref{eq:cv_eic} evaluated at $\Qbar^{*(j)}_{n,-v}$, and $g_n$ is its fit on all $n$ observations.\\
\addlinespace
\textbf{Step 2: Targeting the initial fit.} Fitting~\eqref{eq:linear_fluct} on all $n$ observations gives $\widehat{\epsilon}$ and the targeted regression $\Qbar_n^*=\Qbar_n^0+\widehat{\epsilon}H_{g_n}$.\\
\addlinespace
\textbf{Step 3: Point estimation.} The TMLE is $\widehat{\psi}=n^{-1}\sum_{i=1}^n\{\Qbar_n^*(1,W_i)-\Qbar_n^*(0,W_i)\}$.\\
\addlinespace
\textbf{Step 4: Inference.} Wald confidence intervals and tests are based on the cross-validated influence-curve variance, as in~\eqref{eq:cv_eic}.\\
\addlinespace
\textbf{Cross-validated TMLE.} If the library for the outcome regression contains highly adaptive learners, Steps~1 and~2 are modified as follows \citep{zheng2011cvtmle}.\\
\hangindent=3.25em \hangafter=1 \makebox[3.25em][l]{\textit{Step~1}}Applying (a) and (b) to the observations outside fold~$v$ gives $\Qbar^0_{n,-v}$ and $g_{n,-v}$.\\
\hangindent=3.25em \hangafter=1 \makebox[3.25em][l]{\textit{Step~2}}Fitting~\eqref{eq:linear_fluct} on all $n$ observations, with offset $\Qbar^0_{n,-v(i)}(A_i,W_i)$ and clever covariate $H_{g_{n,-v(i)}}(A_i,W_i)$ for observation~$i$, gives $\widehat{\epsilon}$ and the targeted regressions $\Qbar^*_{n,-v}=\Qbar^0_{n,-v}+\widehat{\epsilon}H_{g_{n,-v}}$.\\
Steps~3 and~4 then proceed as above, with each observation~$i$ evaluated at $\Qbar^*_{n,-v(i)}$ and $g_{n,-v(i)}$.\\
\bottomrule
\end{tabular}
\end{figure}

The library for the treatment mechanism should always include the known $g_0$, so that the Super Learner can retain it when no estimate of the treatment mechanism lowers the cross-validated risk. Its other candidates should be limited to parametric working models, such as logistic regressions of treatment on the covariates considered for adjustment. Estimating the treatment mechanism by maximum likelihood in such a model preserves efficiency \citep{vdlrobins2003}. Penalized regressions or splines can be added if necessary.

To guard against potential overfitting of the nuisance estimators, we use the cross-validated influence-curve variance \citep{balzer2016}. Each observation's influence curve is evaluated at the fits obtained outside its fold, as in~\eqref{eq:cv_eic}, and the sample variance of these evaluations divided by $n$ estimates $\mathrm{Var}(\widehat\psi)$.

\section{Simulation study}\label{sec:simulation}

\subsection{Simulation design and data-generating processes}\label{subsec:sim_design}

The simulation study is calibrated to NPC-002, a 12-month, randomized, double-blind, placebo-controlled trial of arimoclomol for Niemann--Pick disease type~C (NPC), in which $50$ participants were allocated in a $2{:}1$ ratio stratified by baseline miglustat use \citep{mengel2021}. The endpoint considered here is the change from baseline in the rescored four-domain NPC Clinical Severity Scale (R4DNPCCSS), the primary endpoint of the resubmitted new drug application \citep{fda2024briefing,mengel2025}.

Each simulated trial comprises $n=50$ participants. The five baseline covariates, age ($W_1$), sex ($W_2$), years since first symptom ($W_3$), baseline miglustat use ($W_4$), and baseline R4DNPCCSS score ($W_5$), are generated independently as
\begin{equation}\label{eq:sim_covariates}
\begin{aligned}
W_1 &\sim 2+\mathrm{Binom}(17,\,0.52), & W_2 &\sim \mathrm{Bern}(0.53),\\
W_3 &\sim \mathrm{Gamma}(3.2,\,2.4), & W_4 &\sim \mathrm{Bern}(0.8),\\
W_5 &\sim \mathrm{Uniform}\{0,\dots,20\}.
\end{aligned}
\end{equation}
The covariate distributions are based on the marginal summaries reported for NPC-002 \citep{mengel2021,mengel2025}. As in the trial, participants are randomized to arimoclomol ($A=1$) or placebo ($A=0$) in a $2{:}1$ ratio within each stratum of $W_4$.

The outcome $Y$ is the change in R4DNPCCSS from baseline to month 12. We consider two data-generating processes (DGPs) that differ in how $Y$ depends on the covariates. Under DGP~1, a linear adjustment for the prognostic covariates is efficient:
\begin{equation}\label{eq:dgp1}
Y = 4.964 + 0.2\,W_3 - 2\,W_4 - 0.3\,W_5 + \tau A + \varepsilon,
\end{equation}
where $\varepsilon\sim N(0,\,1.3^2)$. Under DGP~2, years since first symptom and the baseline score have nonlinear effects, so the efficiency of an adjustment also depends on its functional form:
\begin{equation}\label{eq:dgp2}
Y = 0.818 + 2.5\,\expit\!\left(\frac{W_3-6}{1.5}\right) - 2\,W_4 + 0.7\,W_5 - 0.0425\,W_5^2 + \tau A + \varepsilon,
\end{equation}
where $\expit(x)=\{1+\exp(-x)\}^{-1}$. Because the treatment effect is additive, $\psi_0=\tau$. Each DGP is simulated in an effect setting with $\tau=-1.3$, within the range of treatment effects on the R4DNPCCSS estimated for the trial, from about $-1.2$ to $-1.7$ across estimands \citep{fda2024briefing,mengel2025}, and in a null setting with $\tau=0$, giving four scenarios. The R code for the simulation study is publicly available at \url{https://github.com/tzhou2801/P-TMLE_rare_diseases}.

\subsection{Candidate estimators and inference procedures}\label{subsec:sim_estimators}

As shown in Section~\ref{sec:gcomp_tmle}, the unadjusted and ANCOVA estimators are special cases of TMLE, so all five candidate estimators in Table~\ref{tab:candidate_estimators} can be viewed as TMLEs that differ only in how they estimate the outcome regression and the treatment mechanism. ANCOVA($W_4,W_5$) adjusts only for the stratification factor and the baseline score, omitting the prognostic covariate $W_3$. ANCOVA($W_3,W_4,W_5$) is correctly specified under DGP~1 but adjusts for the prognostic covariates with a misspecified functional form under DGP~2.

\begin{table}[ht]
\centering
\begin{threeparttable}
\caption{Candidate estimators.}
\label{tab:candidate_estimators}
\small
\setlength{\tabcolsep}{4pt}
\begin{tabular}{@{}c >{\raggedright\arraybackslash}p{0.22\textwidth} >{\raggedright\arraybackslash}p{0.38\textwidth} >{\raggedright\arraybackslash}p{0.28\textwidth}@{}}
\toprule
No. & Estimator & Outcome regression & Treatment mechanism \\
\midrule
1 & Unadjusted & Linear working model containing $A$ & Known $g_0$ \\
2 & ANCOVA($W_4,W_5$) & Linear working model containing $A$, $W_4$, and $W_5$ & Known $g_0$ \\
3 & ANCOVA($W_3,W_4,W_5$) & Linear working model containing $A$, $W_3$, $W_4$, and $W_5$ & Known $g_0$ \\
4 & TMLE+SL ($g_0$) & Discrete Super Learner over the $32$ linear working models containing $A$ and a subset of $\{W_1,\dots,W_5\}$, and a LASSO regression & Known $g_0$ \\
5 & TMLE+SL ($g_n$) & Same as TMLE+SL ($g_0$) & Discrete Super Learner over the known $g_0$ and logistic LASSO regressions of $A$ on main terms and hinge functions, at several penalty levels \\
\bottomrule
\end{tabular}
\end{threeparttable}
\end{table}

TMLE+SL~($g_0$) estimates the outcome regression with the discrete Super Learner, using the cross-validated $\mathrm{eIC}^2$ risk~\eqref{eq:cv_eic} at the known $g_0$ with $V=25$ folds. Because prior knowledge may not indicate which adjustment set to use, the library contains the linear working models for all $32$ subsets of $\{W_1,\dots,W_5\}$, among them the unadjusted and both ANCOVA working models. We also add a LASSO regression \citep{tibshirani1996} on all five covariates, with the intercept and treatment term left unpenalized. The library includes a correctly specified working model under DGP~1 but not under DGP~2.

TMLE+SL~($g_n$) uses the outcome regression of TMLE+SL~($g_0$) and further estimates the treatment mechanism with the discrete Super Learner (Algorithm~\ref{alg:tmle_sl}). The library for the treatment mechanism comprises the known $g_0$ and logistic LASSO regressions of treatment, at several penalty levels, on the main terms $W_1,\dots,W_5$ and hinge functions of $W_1$, $W_3$, and $W_5$ with knots at their sample medians. The hinge functions are included so that the targeting step can capture nonlinearity in the covariate--outcome relationship that the linear outcome working models cannot represent.

Inference for all estimators uses the cross-validated influence-curve variance. The influence curve of TMLE+SL~($g_n$) also accounts for the estimation of the treatment mechanism \citep{vdlrobins2003}. Confidence intervals and two-sided tests use a Student $t$ reference distribution with $n-2$ degrees of freedom.

\subsection{Results}\label{subsec:sim_results}

Tables~\ref{tab:dgp1} and~\ref{tab:dgp2} report the bias, standard deviation (SD), mean squared error (MSE), confidence interval coverage, power, and Type I error rate of each estimator for DGP~1 and DGP~2. Within each DGP, the effect and null settings share the same covariates, treatment assignments, errors, and cross-validation folds, differing only in the value of $\tau$. Each outcome in the effect setting therefore equals the corresponding outcome in the null setting plus $\tau A$. As a result, bias, SD, MSE, and coverage agree in the two settings to the reported precision and are shown once, whereas power and the Type I error rate are specific to the effect and null settings, respectively. In both DGPs, all five estimators have negligible bias and maintain nominal coverage and Type I error control. In particular, TMLE+SL preserves valid inference despite estimating its nuisance functions data-adaptively.

\begin{figure}[htbp]
\centering
\includegraphics[width=0.8\linewidth]{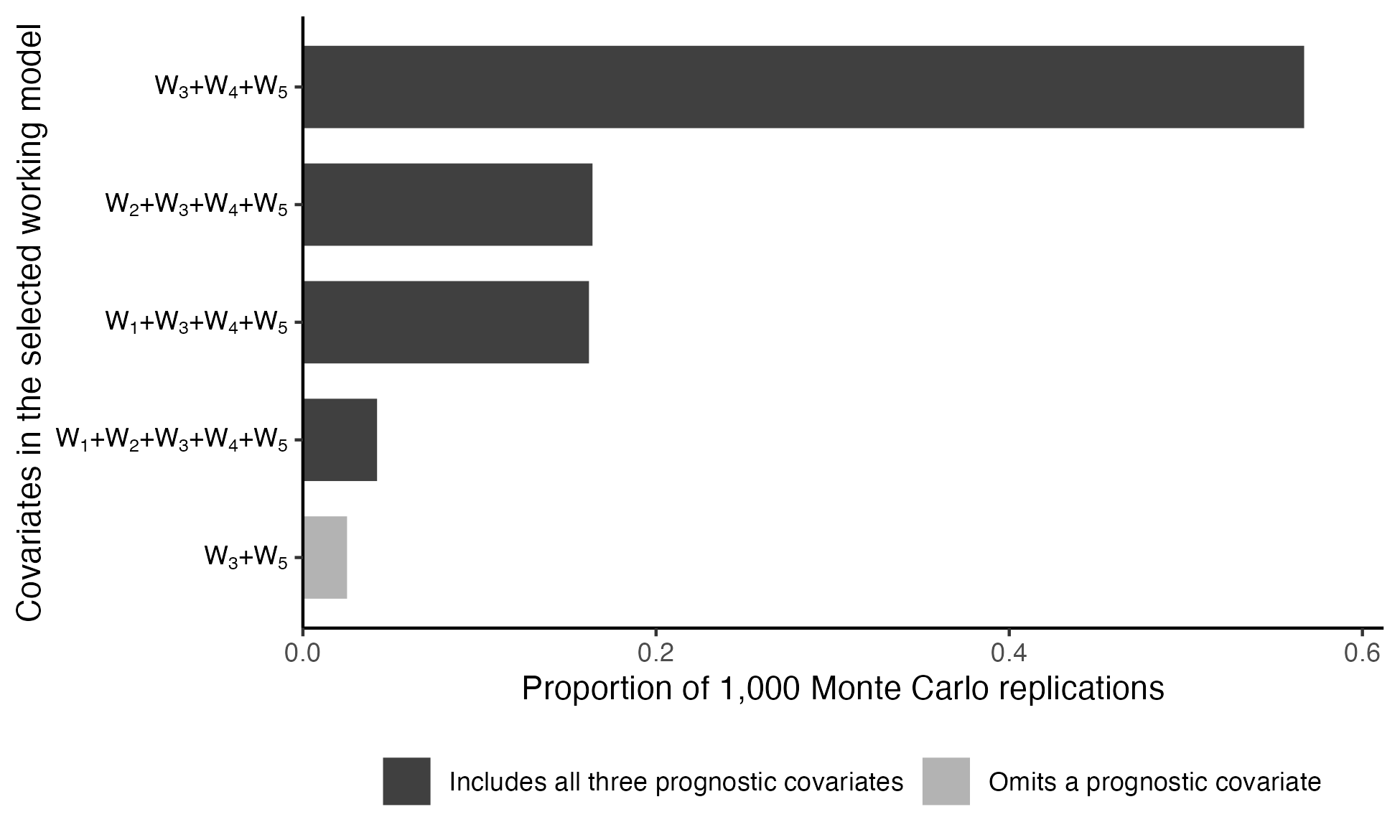}
\caption{Covariates in the outcome working model selected by TMLE+SL~($g_0$) under DGP~1: the proportion of $1{,}000$ Monte Carlo replications in which each covariate set was selected. A LASSO selection is counted by the covariates it retained, and sets selected in fewer than $2\%$ of replications are omitted. Dark bars mark sets that include all three prognostic covariates.}
\label{fig:selection_dgp1}
\end{figure}

Under DGP~1, TMLE+SL~($g_0$) performs comparably to ANCOVA($W_3,W_4,W_5$), the correctly specified and therefore efficient adjustment, in both precision and power. It is also considerably more efficient and powerful than the unadjusted estimator and ANCOVA($W_4,W_5$), which omits the prognostic covariate $W_3$. As Figure~\ref{fig:selection_dgp1} shows, the discrete Super Learner selects a working model that includes all three prognostic covariates in most replications, most often exactly $W_3$, $W_4$, and $W_5$. TMLE+SL thus learns the covariate--outcome relationship from the data rather than requiring it to be known at the design stage, which is often not possible. TMLE+SL~($g_n$) is slightly less efficient and powerful than TMLE+SL~($g_0$). Because the outcome-regression library already contains a correctly specified working model, estimating the treatment mechanism offers no further efficiency gain and only adds finite-sample variability. Even so, TMLE+SL~($g_n$) retains a clear advantage over the unadjusted estimator and ANCOVA($W_4,W_5$).

\begin{table}[htbp]
\centering
\begin{threeparttable}
\caption{Simulation results under DGP~1 ($n=50$; $1{,}000$ Monte Carlo replications).}
\label{tab:dgp1}
\small
\setlength{\tabcolsep}{5pt}
\begin{tabular}{lrrrrrr}
\toprule
Estimator & Bias & SD & MSE & Coverage & Power & \shortstack{Type I\\error rate} \\
\midrule
Unadjusted & $-0.008$ & $0.738$ & $0.545$ & $0.956$ & $0.362$ & $0.044$ \\
ANCOVA($W_4,W_5$) & $-0.009$ & $0.472$ & $0.223$ & $0.958$ & $0.760$ & $0.042$ \\
ANCOVA($W_3,W_4,W_5$) & $0.001$ & $0.389$ & $0.152$ & $0.965$ & $0.878$ & $0.035$ \\
TMLE+SL ($g_0$) & $0.004$ & $0.399$ & $0.159$ & $0.959$ & $0.877$ & $0.041$ \\
TMLE+SL ($g_n$) & $0.001$ & $0.414$ & $0.172$ & $0.964$ & $0.826$ & $0.036$ \\
\bottomrule
\end{tabular}
\begin{tablenotes}[flushleft]
\footnotesize
\item \textit{Note.} Coverage refers to $95\%$ confidence intervals. Power is computed in the effect setting ($\tau=-1.3$) and the Type I error rate in the null setting ($\tau=0$), both for two-sided tests at the $5\%$ level.
\end{tablenotes}
\end{threeparttable}
\end{table}

\begin{table}[htbp]
\centering
\begin{threeparttable}
\caption{Simulation results under DGP~2 ($n=50$; $1{,}000$ Monte Carlo replications).}
\label{tab:dgp2}
\small
\setlength{\tabcolsep}{5pt}
\begin{tabular}{lrrrrrr}
\toprule
Estimator & Bias & SD & MSE & Coverage & Power & \shortstack{Type I\\error rate} \\
\midrule
Unadjusted & $-0.019$ & $0.687$ & $0.472$ & $0.962$ & $0.413$ & $0.038$ \\
ANCOVA($W_4,W_5$) & $-0.022$ & $0.634$ & $0.403$ & $0.959$ & $0.503$ & $0.041$ \\
ANCOVA($W_3,W_4,W_5$) & $-0.012$ & $0.591$ & $0.350$ & $0.966$ & $0.534$ & $0.034$ \\
TMLE+SL ($g_0$) & $-0.038$ & $0.618$ & $0.384$ & $0.946$ & $0.570$ & $0.054$ \\
TMLE+SL ($g_n$) & $-0.022$ & $0.503$ & $0.253$ & $0.954$ & $0.677$ & $0.046$ \\
\bottomrule
\end{tabular}
\begin{tablenotes}[flushleft]
\footnotesize
\item \textit{Note.} Coverage refers to $95\%$ confidence intervals. Power is computed in the effect setting ($\tau=-1.3$) and the Type I error rate in the null setting ($\tau=0$), both for two-sided tests at the $5\%$ level.
\end{tablenotes}
\end{threeparttable}
\end{table}

Under DGP~2, TMLE+SL~($g_0$) achieves efficiency similar to that of ANCOVA($W_3,W_4,W_5$), the most efficient of the three fixed analyses, and greater power than the other two. TMLE+SL~($g_n$) yields the most precise estimates and the highest power of all five estimators. The effects of $W_3$ and $W_5$ are nonlinear in DGP~2, so no linear outcome working model is correctly specified, and the hinge functions in the estimated treatment mechanism allow the targeting step to capture part of the nonlinearity that the selected outcome regression misses. Estimating the treatment mechanism is therefore beneficial when no candidate outcome regression is correctly specified, and costs little otherwise.

\FloatBarrier

\section{Conclusion}\label{sec:conclusion}

We examined TMLE+SL for covariate adjustment in rare-disease RCTs, in which, at the time the analysis must be prespecified, it is rarely known which covariates are prognostic or how they relate to the outcome, so that an analysis restricted to a chosen subset of covariates and a fixed functional form may be suboptimal. TMLE+SL can improve efficiency and power by identifying the prognostic covariates, and learning the form of their relationship to the outcome, from the trial data through a carefully specified library of regression and machine-learning algorithms for the nuisance functions. Estimating the treatment mechanism offers a further opportunity to capture prognostic information that the fitted outcome regression may have missed. The procedure naturally incorporates adjustment methods recommended in the FDA guidance and directly maximizes precision and power. In contrast to naive plug-in estimation with nonparametric fits, the targeting step removes the regularization bias, and influence-curve-based inference remains valid despite the data-adaptive estimation of the nuisance functions, while the procedure as a whole is fully prespecified. We recommend evaluating the operating characteristics of TMLE+SL in simulations that reflect the planned design and sample size, as in Section~\ref{sec:simulation}, to verify that its finite-sample performance is adequate.

For the targeting step, we used the linear fluctuation throughout. Other targeting strategies, such as the logistic fluctuation, can be more stable when positivity is nearly violated \citep{gruber2010bounded}. In an RCT, however, positivity holds by design, so we do not expect the choice of targeting strategy to change our conclusions \citep{xu2026targeting}.

This work has two main limitations. First, we considered a point treatment with a fully observed continuous endpoint. Extensions to longitudinal treatment regimes and to informative censoring are left to future work. Second, our analysis concerned a single primary endpoint. \citet{zhou2026cvtmle} consider related TMLE-based approaches to improving power in trials with multiple endpoints.


\section*{Disclosure Statement}

Tianyue Zhou reports tuition and stipend support from a philanthropic gift from the Novo Nordisk corporation to the University of California, Berkeley, to support the Joint Initiative for Causal Inference. The other authors report no competing interests to declare. The views are those of the author(s) and do not necessarily represent the official views of, nor an endorsement by, FDA, HHS, or the U.S. Government.

\bibliographystyle{plainnat}
\bibliography{refs}

\end{document}